# BRUNO TOUSCHEK AND THE PHYSICS AT FRASCATI AT THE TIME OF ADA AND ADONE (*)

Mario Greco

Department of Mathematics and Physics and INFN,
University of Roma Tre

This is for me a kind of "Amarcord" because I arrived to Frascati in January 1965 joining the ADONE Group, and stayed there for 25 years. AdA had successfully completed his cycle in Orsay [1] and the last publication [2] was at the end of 1964. On the contrary the ADONE Group was in a great turmoil, there was a lot of excitement and a general feeling of sharing a new adventure. The 5th International Conference on High Energy Accelerators was held in Frascati in 1965 and Fernando Amman, the ADONE Group leader, gave an optimistic status report on the construction [3]. Claudio Pellegrini, who was coordinating the theoretical activities of the Group, asked me to face the calculation of the double bremsstrahlung process as a monitor reaction for the luminosity, since the process of single bremsstrahlung, which had studied by G. Altarelli and F. Buccella [4] a year before, coudn't be used due to the large background coming from the bremsstrahlung on the residual gas. That wasn't an easy task, the calculation was not simple, and the two previous estimates of the cross section, by Bayer and Galitsky [5] and Bander [6] differed by a factor of two. At that time this type of calculation required the appropriate handling of Feynman diagrams, traces and integrals totally by hand, and a constant and patient work.

Bruno Touschek at the time was collaborating with Claudio Pellegrini and Enrico Ferlenghi on the instability of the beams [7] and sometimes he was visiting ADONE. In one of those occasions Claudio introduced me to him, also informing him of my work. That was my first personal meeting with him. Bruno was very kind, as usual, and happy that I was proceeding well in my calculation. Indeed he told me that he considered the study of the double bremsstrahlung process quite important and had proposed to Paolo Di Vecchia the study of the soft photon limit as a thesis for his bachelor's degree. Later we joined forces with Paolo and got the final full result using two different approaches [8]. Since that first meeting with Bruno, we started seeing each other regularly in Frascati and also in Rome, where we're living quite closely, and that was the starting of a long collaboration and friendship which lasted many years until 1978, when both of us were visiting Cern and he was hospitalized at the Hopital de la Tour.

(*) Invited talk at the Bruno Touschek Memorial Symposium, December 2-4, 2021.

At the end of 1965 my fellowship was over and Bruno proposed me to join a new theory group he was going to found in Frascati. That took place in a short while by adding together the new young forces of Bruno's students and some newly graduates in Rome. Gian De Franceschi, a mathematician who was already in Frascati with Raoul Gatto before the latter moved to Florence, also joined the group. In more detail, Gian De Franceschi, M.G., Etim Etim, Giulia Pancheri, Paolo Di Vecchia, Giancarlo Rossi, Francesco Drago and Pucci Di Stefano composed the new theory group. The radiative corrections for ADONE experiments were the main problem Bruno had in his mind, because of their importance in the new electron-positron collider. Indeed he had already given a number of theses to his students, namely "Proposal for the administration of radiative corrections" to Etim Etim, "The double bremsstrahlung process in the soft photon approximation" to Paolo di Vecchia and "Application of the Block-Nordsieck theorem to radiative corrections in Adone experiments" to Giancarlo Rossi. To summarize the further theoretical efforts in that direction, two main research lines had been followed.

First, the infrared corrections to be applied in an electron-positron collider experiment are obtained with the help of the Bloch-Nordsieck theorem, using a statistical approach to define the probability for the four-momentum to be carried away by the electromagnetic radiation [9]. Alternatively, from a field theoretical point of view, a new finite S-Matrix is defined using a realistic definition of initial and final states, by "dressing" the charged particle's states with a phase containing the electromagnetic operator in the exponential, in order to create an undetermined number of soft photons. The new S-Matrix was explicitly shown to be equivalent to all orders in $\alpha$ to the conventional perturbative result [10]. In other words this approach corresponds to the introduction of the coherent states in QED. It is extraordinary that both approaches led exactly to the same result for the soft radiative effect, namely the observed cross section can be written as

$$d\sigma = [\gamma^\beta \Gamma(1+\beta)]^{-1} (\Delta\omega/E)^\beta d\sigma_E$$

where 2E is the total c.m. energy, $(\Delta\omega/E)$ is the relative energy resolution of the experiment, $\gamma$ is the Euler constant, $d\sigma_E$ differs from the lowest cross section $d\sigma_0$ by finite terms of o ($\alpha$), and $\beta$ is the famous Bond-factor, so named by Bruno because its numerical value at ADONE was 0.07, and more generally $\beta = 4[\log (2E/m) - 1/2]/\pi$.

The coherent states approach played a major role later in the description of the radiative effects in case of the production of the J/Ψ and of the Z boson, as we'll discuss later. Also that was first extended to QCD in the late '70 [11] and studied further [12-13]. Many and important QCD results concerning exponentiation, resummation formulae, K-factors, transverse momentum distributions of DY pairs, W/Z and H production, have their roots in Bruno Touschek ideas on the exponentiation and resummation formulae in QED. We give here a reference list [14] of the papers that were written later in the Frascati-Rome area and certainly inspired by his ideas.

Coming back to the late '60, a strong shock was inflicted to the physics community due to an apparent violation of QED reported by a Harvard group, R.B. Blumenthal et al., in the wide-angle production of electron-positron pairs [15]. New experiments immediately took place everywhere, and also Carlo Bernardini at Frascati started an experiment of wide-angle bremsstrahlung. On the theory side Bruno Touschek suggested a simple method of modifying QED by introducing ee$\gamma\gamma$ vertices in addition to the usual minimal ee$\gamma$ interaction, and asked me to study the possible constraints coming from the known effects and experiments. When the draft of paper was ready, the news arrived of the confirmation of the validity of QED, and the paper got unpublished [16]. The new experiments were summarised by C. Bernardini [17] and reported in Fig.1. It's amusing to notice that the first and last authors in WAEP experiments [18] are B. Richter and S.C.C. Ting who will share in a few years the discovery of the J/$\Psi$.

Let's discuss now the theoretical framework and the expectations concerning ADONE and the experimental results. At the time the Vector Meson Dominance (VMD) model of J.J. Sakurai [19] was quite successful in describing the e.m. interaction of hadrons as being mediated by the vector mesons $\rho$, $\omega$ and $\varphi$. That led T.D. Lee, N. Kroll and B. Zumino to try to give a field theoretical approach to VMD [20]. In this framework the total hadronic annihilation cross section was expected to behave at large s as

$$\sigma(s) \approx (1/s)^2$$

However departures from the simple VMD model were observed in some radiative decays of mesons and the possible existence of new vector mesons was suggested by A. Bramon and myself [21], as also predicted by dual resonance models and the Veneziano model [22]. On the other hand the results of Deep Inelastic Scattering (DIS) experiments at SLAC, with the idea of Bjorken scaling and the Feynman parton model were naturally leading to

$$\sigma(s) \approx (1/s)$$

and indeed N. Cabibbo, M. Testa and G. Parisi [23] suggested that the ratio R of the hadronic to the pointlike cross section would asymptotically behave as

$$R = \sigma_{had}(s) / \sigma_{\mu\mu}(s) \longrightarrow \Sigma_i Q_i^2$$

where the sum extends to all spin 1/2 elementary constituents, neglecting scalars.

As it's well known, the results of all experiments, namely the MEA Group [24],

the γγ Group [25], the μπ Group [26] and the Bologna-CERN-Frascati Collaboration [27], showed a clear evidence of a large multihadron production with R ≈ 2, pointing to the coloured quark model. On the other hand they also indicated evidence for a new vector meson ρ' (1.6) with a dominant decay in four charged pions, which had been suggested by A. Bramon and myself [28]. The experimental data are shown in Figs. [2-3], taken from a review paper of C. Bernardini and L. Paoluzi [29].

The ADONE results together with the request of scaling, both in DIS and e+e- annihilation, and the Veneziano's duality idea led us to propose a new scheme where the asymptotic scaling is reached through the low energy resonances mediating the asymptotic behaviour [30]. Thus the value of R is also connected to the low energy resonances's couplings, and in the 3 coloured quark model, led us to the prediction R ≈ 2.4. This scheme - named duality in e+e- annihilation - was immediately shared by J. J. Sakurai [31]. Later J. Bell and collaborators also studied a potential model where the bound states could be solved analytically and verified this idea of duality [32].In addition a set of e+e- duality sum rules was derived from the canonical trace anomaly of the energy momentum tensor by E. Etim and myself [33], much earlier than the Russian sum rules of M. A. Shifman et al. [34]. The lowest order sum rules gives

$$\int_{s_0}^{\bar{s}} ds \left( Im\Pi(s) - \frac{\alpha R}{3} \right) = 0,$$

where ImΠ (s) = s σ$_{had}$(s) / 4πα, and clearly it relates the asymptotic value of R to the low energy behaviour. One has to stress here that QCD wasn't there yet at that time. Fifty years later, by comparing now the value of R from the Particle Data Group with all the experimental information, as shown in Fig. [4], with the theoretical prediction of QCD with o ($α_s$), o ($α_s^2$) and o ($α_s^3$) corrections included - as indicated by the continuous red line - one easily concludes the duality is very well satisfied. The average value of R in particular, in the ADONE region, is about 2.4 as it was also confirmed by the SPEAR data at the c.m. energy just below the J/Ψ.

Let's consider in detail the J/Ψ discovery, or what was called the November Revolution, from a Frascati point of perspective. As it's well known, on November 11th 1974, B. Richter and S.C.C. Ting jointly announced in Stanford the discovery of the J/Ψ both at SLAC and at Brookhaven [35,36]. I had the terrific chance of arriving at SLAC the day after, with an invitation by Sid Drell to give a seminar on our duality works, on the way for a visit of a few weeks to Mexico City. Sid had been on a sabbatical leave the year before at Frascati and Rome, so we knew each other pretty well. There was a great excitement in the theory discussion room and once I was informed of the details of the discovery I realized immediately that the J/Ψ could be seen possibly also at ADONE. I asked Sid to let me call Frascati, and from his confidential office - he was scientific advisor of the President of United States - I gave to Giorgio Bellettini, the director of the Laboratory, the exact position of the J/Ψ. The night after, the resonance was also observed at Frascati. Giorgio Salvini communicated the results to the Phys. Rev. Letters over the telephone and the paper was published [37] in the same issue of the American results.

As far as the theoretical interpretation of the J/Ψ, hundreds of papers had been published on the argument, as it's well known. In a recent review article on this subject, Alvaro De Rujula has reported [38] the papers published on the first issue of Phys. Rev. Letters after the discovery, as shown in Fig. [5]. Of course only two of them, both from Harvard, had the right interpretation: by T. Applequist and H.D. Politzer [39], who related the reason for the very narrow width of the J/Ψ to the asymptotic freedom of QCD just discovered, and by A. De Rujula and S.L. Glashow [40] because of the GIM mechanism and charm suggested earlier [41]. On the other hand Alvaro is also quoting two papers published on Lett. Nuovo Cimento, by G. Altarelli, N. Cabibbo, L. Maiani, R. Petronzio and G. Parisi [42], who had the wrong interpretation in favour of the weak boson, and C. Dominguez and myself [43], written in Mexico City a few days after I had left Stanford, who also had the right interpretation in favour of charm. In more detail, as soon as I found the news on a local newspaper of the subsequent discovery of the Ψ', by using the duality ideas discussed above, we arrived at the conclusion that the new series of resonances was indeed composed by c-c̄ pairs. An enjoyable note concerning this paper came fourty years later. On December 2013, S.C.C. Ting was invited to Frascati for a Bruno Touschek Memorial Lecture and commenting the J/Ψ saga, he said that our paper was the first one to give the right interpretation. I was very surprised because I had never compared the exact dates of the three papers, and our original preprint had been lost and not easily available. However, searching virtually at the CERN library archives I indeed found that our paper was preceding by one day that one of Applequist and Politzer and by one week the other by De Rujula and Glashow.

The problem of the radiative corrections to the J/Ψ line-shape, in virtue of the very narrow width, showed the crucial role played by the theoretical ideas of the early times on the infrared behaviour of QED, namely the exponentiation results and the approach of the coherent states. The detailed analysis by G. Pancheri, Y. Srivastava and myself [44], showed that the main infrared correction factor was of the type

$$C_{infra} \approx (\Gamma/M)^{\beta}$$

where $\Gamma$ and $M$ are the width and the mass of the resonance respectively, and $\beta$ the Bond-factor. The detailed result of this analysis, compared with the SLAC and Frascati data, is shown in Fig. [6]. When we showed our result to Bruno Touschek, he immediately commented that the experimental errors of the Frascati data had been overestimated. At this point I should add that the SLAC analysis of their data had been based on a paper by D.R. Yennie [45] that contained a wrong dependence on the width Γ and the parameter σ of the Gaussian energy distribution of the beams, with a resulting difference with respect to our analysis on the leptonic width of the J/Ψ. It was only in 1987, in occasion of the first La Thuile meeting, that I convinced Burt Richter to update the SLAC radiative corrections codes with the right formulae, in perspective of the coming data on the Z boson physics at SLC. Indeed the re-analysis of all charm data at SLAC caused a change of many properties of the charm particles in the Particle Data Group in 1988.

The above treatment of the radiative corrections for the J/Ψ production was extended a few years later to study the radiative effects in the case of Z production at LEP/SLC [46]. Our work was the first study to all orders in the infrared corrections, with a complete evaluation of all finite terms of o (α), and at the base of the later analyses of the experiments. Very recently, within the general discussion on the possibility of constructing a muon collider Higgs factory to study with great care on resonance the properties of the H, the line-shape has been studied [47] in the same way, as in the old times. As a result we have shown that the radiative effects put very stringent bounds on the energy spread of the beams, and make this project very tough.

To conclude, from AdA/ADONE to LEP/SLC and the future linear and/or circular colliders, the seminal idea of Bruno Touschek has contributed with so many discoveries to the assessment and the progress of the Standard Model. This certainly constitutes his main legacy. In addition he has given many ideas in theoretical physics, from QED to other aspects of the Standard Model, and that is also an important legacy to us.

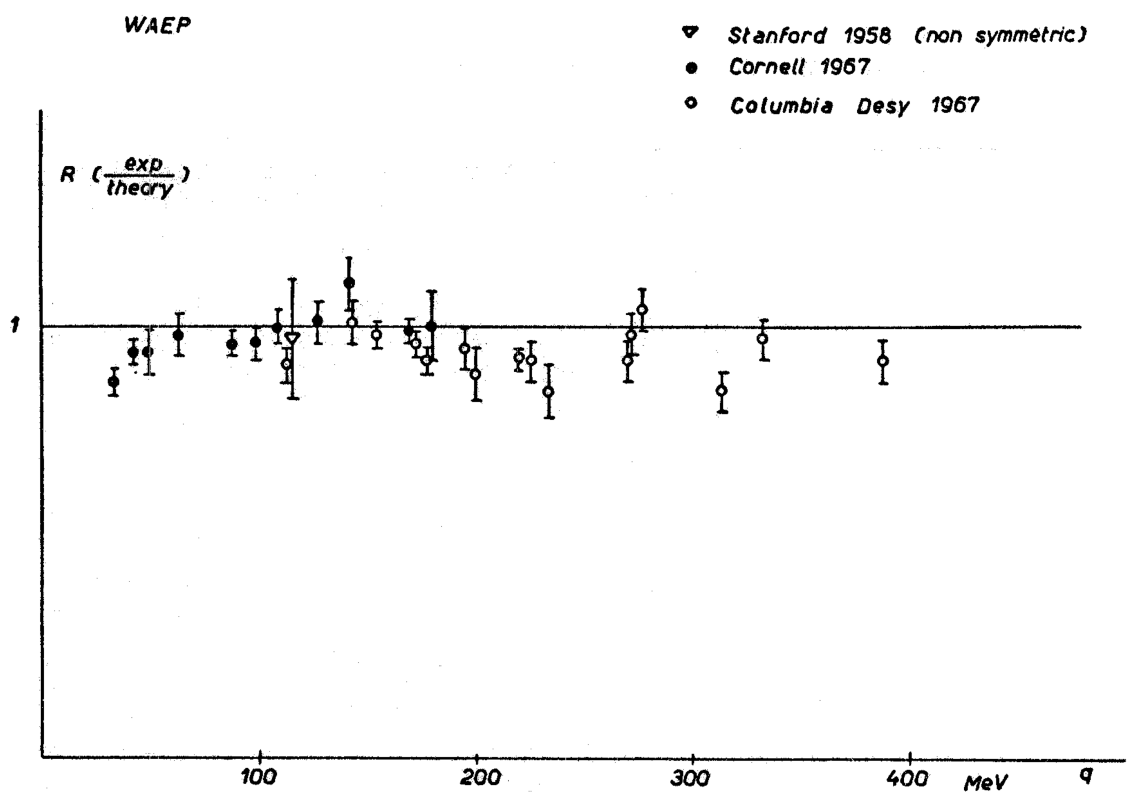

Fig. 1

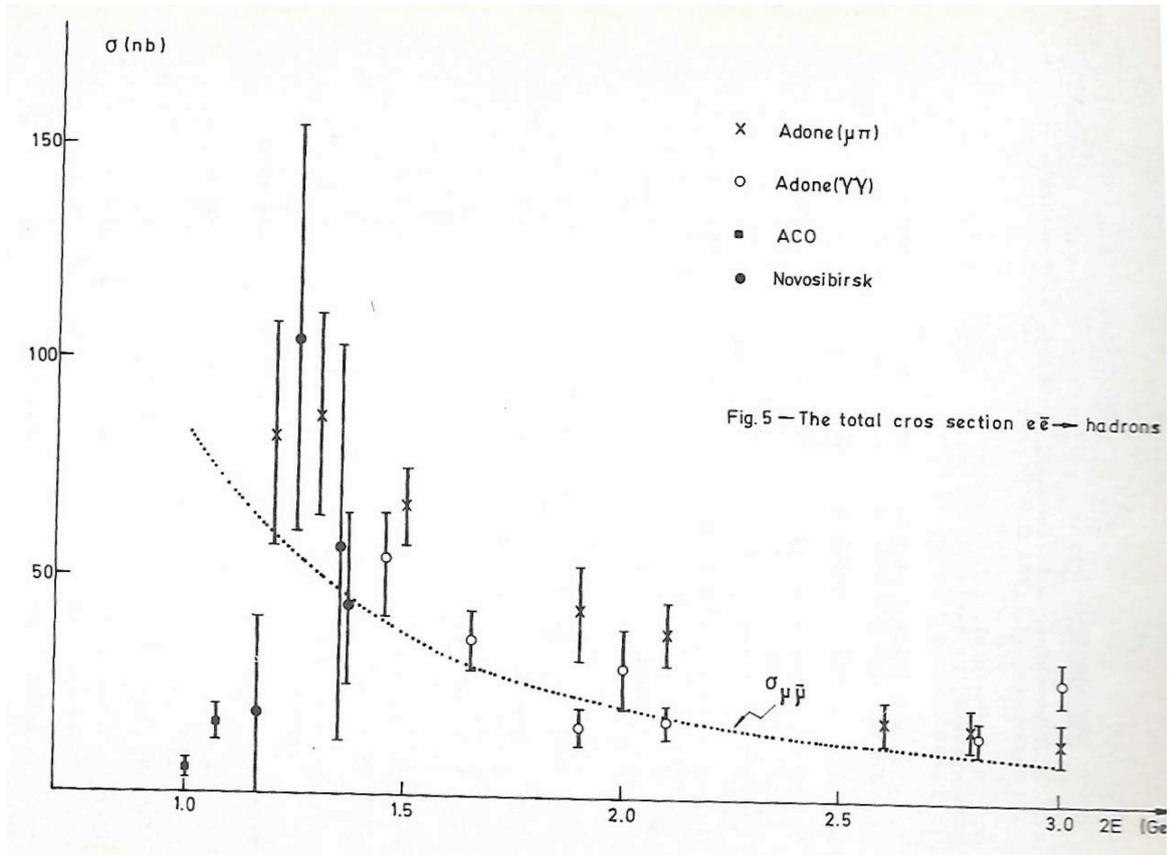

Fig. 2

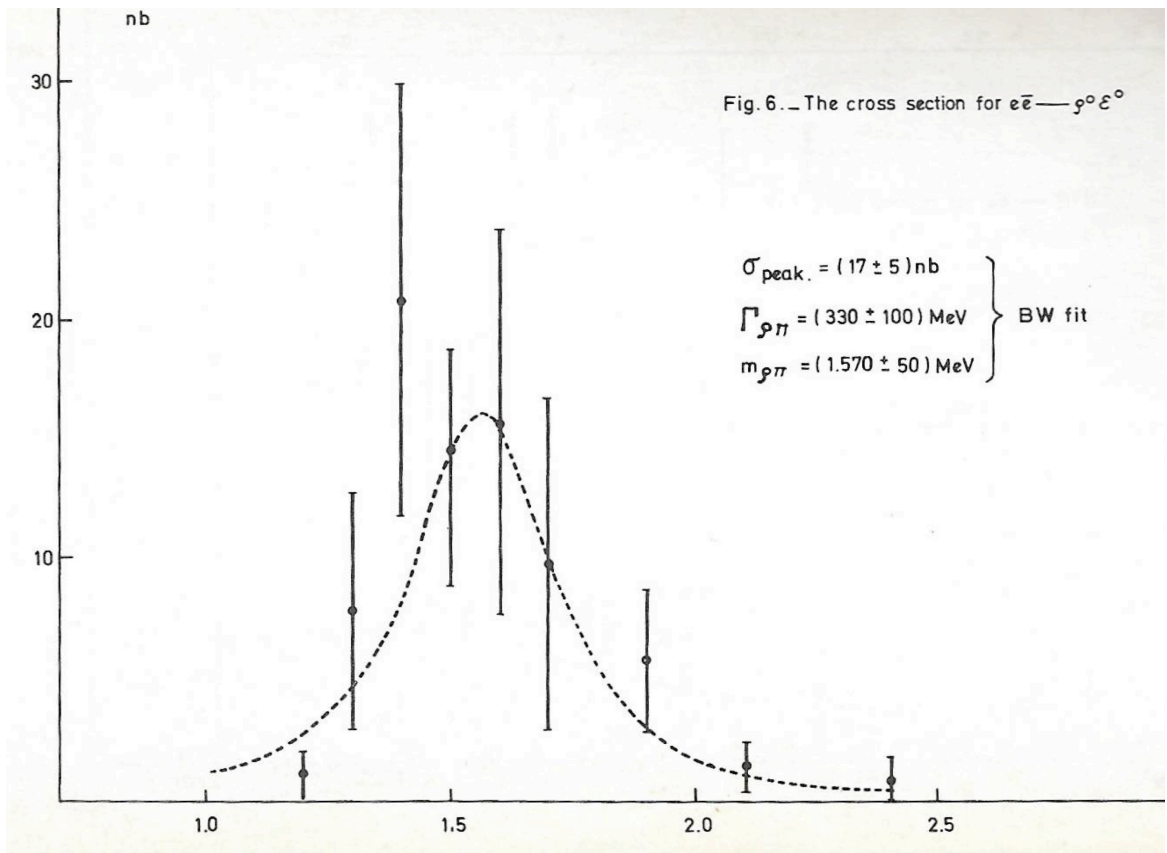

Fig. 3

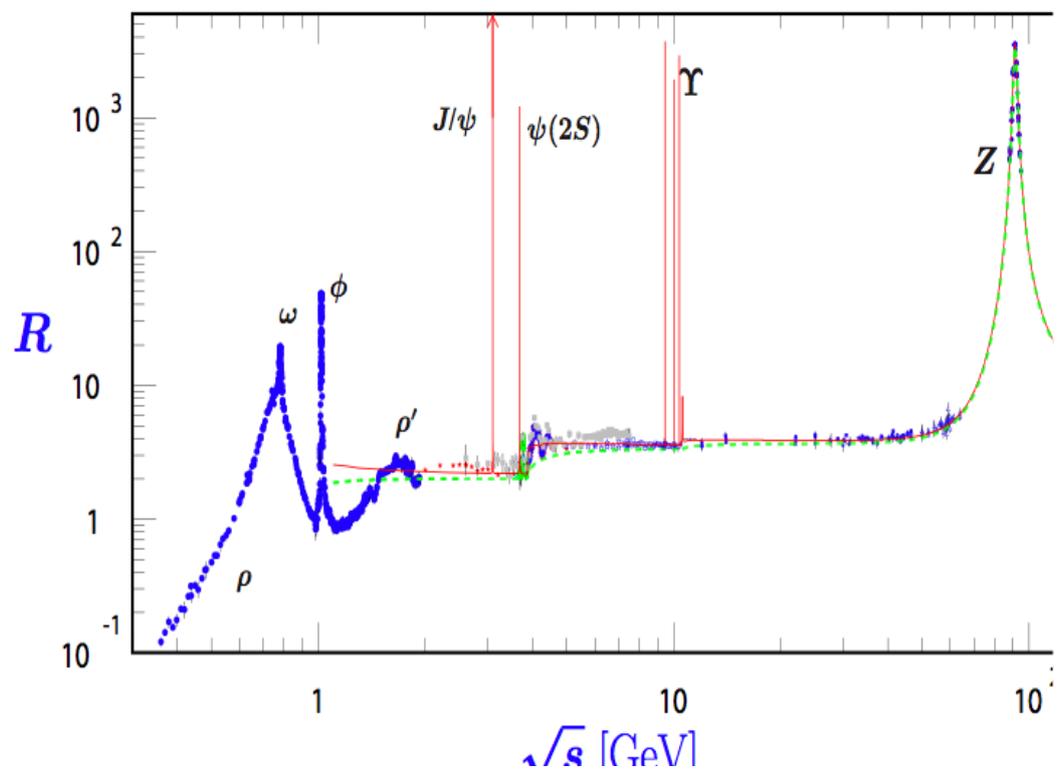

Fig. 4

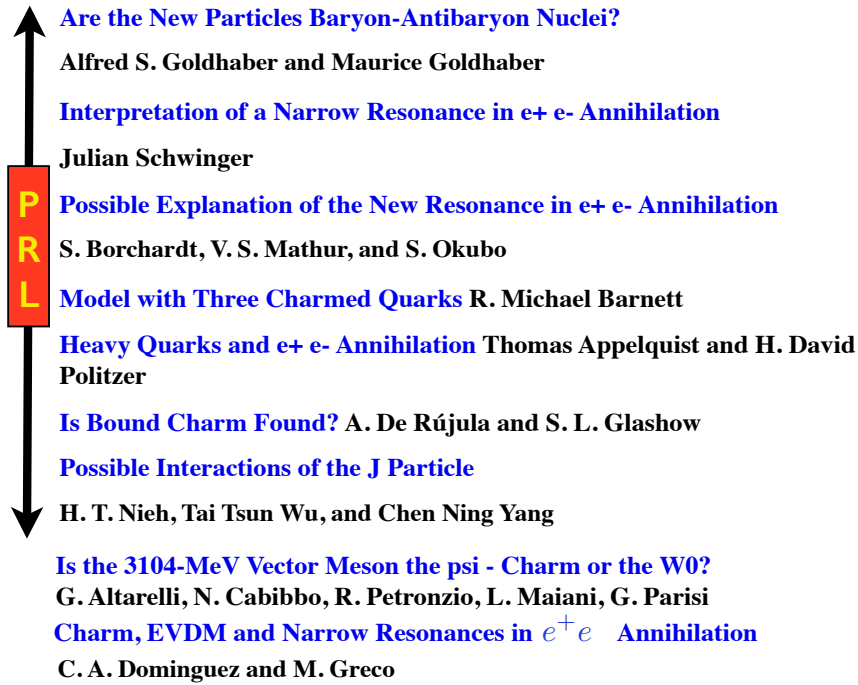

Fig. 15. Immediate interpretations of the $J/\psi$, with their titles. PRL is Phys. Rev. Lett. **4**, Jan. 6th, 1975. The last two papers[88,89] are in Lett. Nuovo Cim.

Fig. 5

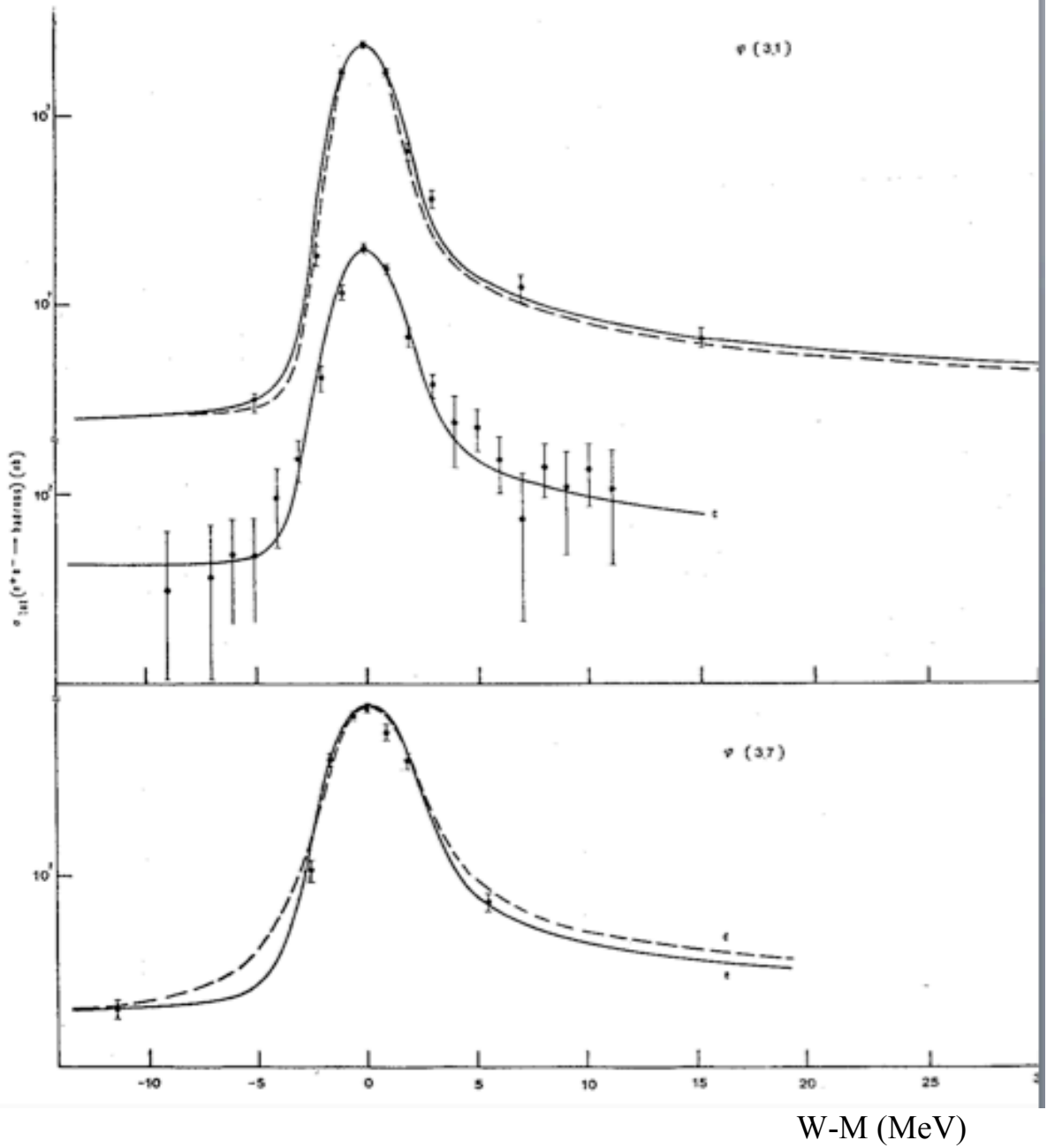

Fig. 6 - Data from SPEAR and ADONE